\documentclass[twocolumn,showpacs,preprintnumbers,amsmath,amssymb]{revtex4}

\usepackage{graphicx}
\usepackage{dcolumn}
\usepackage{bm}

\begin{document}

\preprint{APS/123-QED}

\title{Field-enhanced quantum fluctuation in $S$=1/2 frustrated square lattice}

\author{H. Yamaguchi$^1$, Y. Sasaki$^1$, T. Okubo$^2$, M. Yoshida$^3$, T. Kida$^4$, M. Hagiwara$^4$, Y. Kono$^3$, S. Kittaka$^3$, T. Sakakibara$^3$, M. Takigawa$^3$, Y. Iwasaki$^1$, and Y. Hosokoshi$^1$}
\affiliation{
$^1$Department of Physical Science, Osaka Prefecture University, Osaka 599-8531, Japan \\ 
$^2$Department of Physics, the University of Tokyo,Tokyo 113-0033, Japan\\
$^3$Institute for Solid State Physics, the University of Tokyo, Chiba 277-8581, Japan \\
$^4$Center for Advanced High Magnetic Field Science (AHMF), Graduate School of Science, Osaka University, Osaka 560-0043, Japan \\}

Second institution and/or address\\
This line break forced

\date{\today}

\begin{abstract}
We present a new model compound with the $S$ = 1/2 frustrated square lattice composed of the charge-transfer salt ($o$-MePy-V)PF$_6$.
$Ab$ $initio$ calculations indicate the formation of an $S$ = 1/2 square lattice, in which six types of nearest-neighbor ferromagnetic- and antiferromagnetic interactions cause frustration.
By applying a magnetic field, we observe an unusually gradual increase of magnetization and a subsequent 1/2-plateau-like behaviour.
A numerical analysis using the tensor network method qualitatively demonstrates such behaviors and suggests a collinear ordered state and a field-enhanced quantum fluctuation.  
Furthermore, the local magnetization and $T_1^{-1}$ probed by nuclear magnetic resonance measurements support these findings.
\end{abstract}

\pacs{75.10.Jm, 
}

\maketitle
\section{INTRODUCTION}
One of the key focus areas in condensed matter physics is the search for novel quantum phenomena caused by strong quantum fluctuations. 
In this regard, one-dimensional (1D) spin systems have been extensively studied both theoretically and experimentally, and it has been confirmed that exotic quantum states, such as a Tomonaga-Luttinger liquid~\cite{tomonaga,luttinger} and a Haldane state~\cite{haldane}, can be realized in such systems instead of conventional magnetic orders owing to strong quantum fluctuations.
In the case of two-dimensional (2D) spin systems, since the discovery of high-temperature superconductivity in layered cuprates in 1986, considerable attention has been devoted to the quantum state of their parent spin system, i.e., the $S$ = 1/2 square lattice~\cite{cu_super,cu_nature}, and a unique quantum spin state in the form of a resonating-valence-bond has been discussed~\cite{RVB_square, RVB_rev}.  
Although extensive studies on the $S$ = 1/2 square lattice antiferromagnet have established that its ground state exhibits a conventional N$\rm{\acute{e}}$el order, it has been revealed that the quantum fluctuations reduce the magnetic moment per site by approximately 40 ${\%}$ with respect to the classical value, and cause renormalization of the spin wave energy~\cite{square, re_spinwave}.

In $S$ = 1/2 quantum spin systems, novel quantum phases are expected to appear with the introduction of frustration through competing exchange interactions that cannot be simultaneously satisfied~\cite{RVB, L_balents}.
For the square lattice, the $S$ = 1/2 $J_{1}-J_{2}$ model, in which the nearest-neighbour interaction $J_{1}$ and next-nearest-neighbour interaction $J_{2}$ cause a frustration, is known as a prototype of a frustrated model.
An interesting phase diagram with unusual ground states is theoretically predicted as a function of the ratio $J_{2}/J_{1}$.
Furthermore, quantum phases have been predicted to exhibit several novel ground states such as valence-bond crystal, spin-nematic phase, and spin liquids with or without a spin gap, though the true ground states have not been experimentally verified yet~\cite{VBC2,nematic, gap1,gapless1,gapless2}.
Theoretical investigations in magnetic fields also indicate the appearance of several field-induced quantum phases~\cite{morita22, morita24}.  
Considering the variety of predicted quantum phases in the $S$=1/2 $J_{1}-J_{2}$ model due to the interplay of quantum fluctuation and frustration, different types of frustrated interactions in the $S$=1/2 square lattice are also expected to give rise to exotic quantum phenomena.
In particular, a frustrated square lattice formed by only nearest-neighbour interactions~\cite{TCNQ_square} causes much stronger quantum fluctuations owing to its relatively small coordination number (i.e,4), which is the same as that of the kagome lattice~\cite{kagome_science, herber, hotta}.

In this letter, we present a new model compound of the $S$ = 1/2 frustrated square lattice with only nearest-neighbour interactions.
We successfully synthesized single crystals of the verdazyl-based charge-transfer salt ($o$-MePy-V)PF$_6$ [$o$-MePy-V = 3-(2-methylpyridyl)-1,5-diphenylverdazyl]. 
$Ab$ $initio$ molecular orbital (MO) calculations indicate the formation of an unprecedented $S$ = 1/2 square lattice, in which six types of nearest-neighbour ferromagnetic (FM) and antiferromagnetic (AFM) interactions cause frustration.
We observe an unconventional gradual increase of the magnetization curve and a subsequent 1/2-plateau-like behavior.
A numerical analysis using the tensor network method qualitatively demonstrates unconventional magnetic behaviors accompanied by field-enhanced quantum fluctuations.  
Furthermore, the local magnetization and $T_1^{-1}$ probed by nuclear magnetic resonance (NMR) measurements show behaviors consistent with those expected from the numerical analysis.

\section{EXPERIMENTAL AND NUMERICAL METHOD}
The crystal structure was determined on the basis of intensity data collected using a Rigaku AFC-8R Mercury CCD RA-Micro7 diffractometer with Japan Thermal Engineering XR-HR10K. 
The magnetizations were measured using a commercial SQUID magnetometer (MPMS-XL, Quantum Design) and a capacitive Faraday magnetometer down to about 70 mK.
High-field magnetization measurement in pulsed magnetic fields of up to approximately 50 T was conducted using a non-destructive pulse magnet.
The experimental result was corrected for the diamagnetic contribution calculated using the Pascal method.
The specific heat was measured by a standard adiabatic heat-pulse method down to about 0.3 K.
All above experiments were performed using single crystals with typical dimensions of 1.0$\rm{\times}$1.0$\rm{\times}$0.5 mm$^3$, which are stable at experimental temperature.

NMR measurements were performed using a single crystal, and an external magnetic field was applied along the $b$ axis. 
The $^{31}$P-NMR spectra were obtained with a fixed magnetic field by summing the Fourier transform of the spin-echo signal at equally spaced rf frequencies. 
We determined 1/$T_{1}$ by fitting the spin-echo intensity $M$($t$) as a function of the time $t$ after the inversion pulse to the stretched exponential recovery function $M$($t$) = $M_{\rm{eq}}$ $-$ $M_{0}$exp${\{}$$-$($t$/$T_1$)$^{\beta}{\}}$,
where $M_{\rm{eq}}$ is the intensity at thermal equilibrium and ${\beta}$ is the stretch exponent that provides a measure of inhomogeneous distribution of 1/$T_{1}$.

$Ab$ $initio$ MO calculations were performed using the UB3LYP method in the Gaussian 09 program package.
The basis set is 6-31G. 
For the estimation of intermolecular magnetic interaction, we applied our evaluation scheme that have been studied previously~\cite{MOcal}.

\begin{figure}[t]
\begin{center}
\includegraphics[width=18pc]{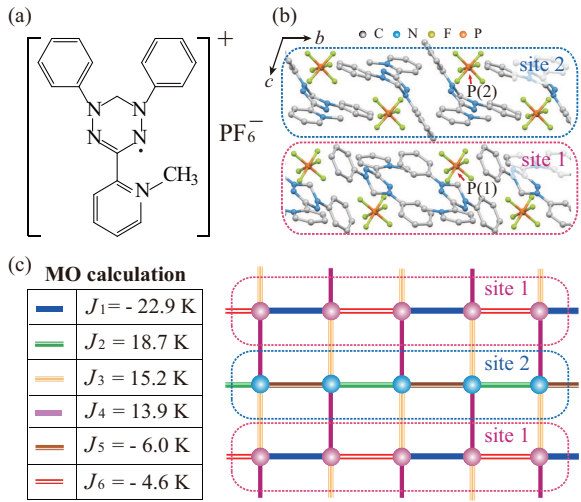}
\caption{(color online) (a) Molecular structure of ($o$-MePy-V)PF$_6$. (b) Crystal structure of ($o$-MePy-V)PF$_6$ viewed  parallel to the $a$ axis. Hydrogen atoms are omitted for clarity. The broken lines indicate the crystallographically independent sites, site 1 and site 2, along the $b$ axis. (c) $S$ =1/2 square lattice composed of $J_{i}$ ($i=1-6$). The left table lists the values of the exchange interactions evaluated from the $ab$ $initio$ MO calculation.}\label{f1}
\end{center}
\end{figure}

In order to calculate the ground state of an effective model, we conducted tensor network method which can treat infinite-size system directly. 
We used the tensor network ansatz so called infinite Tensor Product State (iTPS)\cite{iTPS1,iTPS2} or infinite Projected Entangled Pair State (iPEPS) \cite{iPEPS1,iPEPS2,iPEPS3}. 
Recently the iTPS has been applied to frustrated spin systems including the Kagome lattice Heisenberg model \cite{Kagome_TN1,Kagome_TN2,Kagome_TN3}, the Shastry-Sutherland lattice Heisenberg model \cite{SS_TN1,SS_TN2} and
models with Kitaev interaction\cite{Kitaev_TN1,Kitaev_TN2}. 
We assumed infinitely repeated $4\times 2 $ unit-cell structure forming the square lattice network as shown in Fig.\ref{f1} with the bond dimensions up to $D = 6$.  
For the optimization of the tensors, we used the imaginary-time evolutions with so called simple update technique
\cite{ITE,ITE_memo} and the environment was calculated by the corner transfer matrix method \cite{CTM1,CTM2,CTM3,CTM4,CTM5}.

\section{RESULTS AND DISCUSSION}
\subsection{Crystal structure and magnetic model}
Figure 1(a) shows the molecular structure of ($o$-MePy-V)PF$_6$.
The crystallographic parameters at room temperature and 25 K are summarized in Table I, and there was no indication of a structural phase transition.
Because this investigation focused on the low-temperature magnetic properties, the crystallographic data at 25 K are discussed hereafter.
The crystallographic parameters at 25 K were as follows: triclinic, space group $P\bar{1}$, $a$ =  10.685(13) $\rm{\AA}$, $b$ = 11.537(14) $\rm{\AA}$, $c$ = 15.79(2) $\rm{\AA}$, $V$ = 1899(4) $\rm{\AA}^3$, and $Z$ = 4. 
The crystals contain two crystallographically independent $o$-MePy-V molecules.
Each molecule forms a 1D structure related by inversion symmetry along the $b$-axis, as shown in Fig. 1(b). 
We performed MO calculations in order to evaluate the exchange interaction between $S$ = 1/2 spins on $o$-MePy-V molecules and found six types of dominant interactions.
They are evaluated as $J_{1}/k_{\rm{B}}$ = $-22.9$ K, $J_{2}/k_{\rm{B}}$
= 18.7 K, $J_{3}/k_{\rm{B}}$ = 15.2 K, $J_{4}/k_{\rm{B}}$ = 13.9 K,
$J_{5}/k_{\rm{B}}$ = $-6.0$ K, and $J_{6}/k_{\rm{B}}$ = $-4.6$ K, which
are defined in the Heisenberg spin Hamiltonian given by $\mathcal {H} =
J_{n}{\sum^{}_{\langle i,j \rangle}}\textbf{{\textit
S}}_{i}{\cdot}\textbf{{\textit S}}_{j}$, where $\sum_{ \langle i,j \rangle}$ denotes the sum over the neighboring spin pairs.
The molecular pairs related to $J_{1}$, $J_{2}$, $J_{5}$, and $J_{6}$ correspond to those forming the 1D structure along the $b$-axis.
Thus, two spin chains composed of $J_{1}$-$J_{6}$ (site 1) and $J_{2}$-$J_{5}$ (site 2) exist along the $b$-axis, and they are alternatingly coupled by AFM interactions $J_{3}$ and $J_{4}$, resulting in the formation of a 2D frustrated square lattice in (10$\bar{1}$) plane, as shown in Fig. 1(c).   
The PF$_6$ anions act as a spacer between 2D square planes and enhance the 2D character of the present model.

\begin{table}
\caption{Crystallographic data for ($o$-MePy-V)PF$_6$.}
\label{t1}
\begin{center}
\begin{tabular}{ccc}
\hline
\hline 
Formula & \multicolumn{2}{c}{C$_{20}$H$_{19}$F$_{6}$N$_{5}$P}\\
Crystal system & \multicolumn{2}{c}{Triclinic}\\
Space group & \multicolumn{2}{c}{$P$$\Bar{1}$}\\
Temperature (K) & RT & 25(2)\\
Wavelength ($\rm{\AA}$) & \multicolumn{2}{c}{0.7107} \\
$a (\rm{\AA}$) &  10.919(4) & 10.685(13) \\
$b (\rm{\AA}$) &  11.814(5) & 11.537(14) \\
$c (\rm{\AA}$) &  16.222(7) & 15.79(2)\\
$\alpha$ (degrees) &  96.645(6)  & 93.72(3)\\
$\beta$ (degrees) &  102.921(6)  & 101.565(14)\\
$\gamma$ (degrees) &  93.669(4)  & 92.84(3)\\
$V$ ($\rm{\AA}^3$) & 2017.1(14) &  1899(4) \\
$Z$ & \multicolumn{2}{c}{4} \\
$D_{\rm{calc}}$ (g cm$^{-3}$) & 1.562 & 1.659\\
Total reflections & 6640 & 5811\\
Reflection used & 3551 & 3627\\
Parameters refined & \multicolumn{2}{c}{531}\\
$R$ [$I>2\sigma(I)$] & 0.0539 & 0.0846\\
$R_w$ [$I>2\sigma(I)$] & 0.1134 & 0.1961\\
Goodness of fit & 0.959 & 1.171\\
CCDC &  1573180 & 1573181 \\
\hline
\hline
\end{tabular}
\end{center}
\end{table}

\subsection{Specific heat}
The specific heat $C$/$T$ at zero-field exhibits a sharp peak at $T_{N}$=1.7 K, which is associated with a phase transition to the ordered state, as shown in Fig. 2(a).
The magnetic contributions are expected to be dominant in the low-temperature regions considered here~\cite{3Cl4FV, 2Cl6FV, b26Cl2V, pBrV}, and nuclear Schottky contributions are subtracted assuming estimation from the nuclear spins, 2.239${\times}$ $10^{-4}$$H^2$/$T^2$.
In the low-temperature regions below approximately 1 K, $C$/$T$ shows a clear $T$-linear behaviour, which corresponds to a $T^2$ dependence of the specific heat attributed to a linear dispersive mode in the 2D AFM system.
The entropy $S$, which can be obtained through the integration of $C/T$, demonstrates that less than 1/3rd of the total magnetic entropy of 5.76 ($R$ln2) is associated with the phase transition, while the residual entropy is almost entirely consumed above $T_{N}$, as shown in the inset of Fig. 2(a). 
This behaviour also indicates that the present spin system has a good 2D character, yielding a sufficient development of short-range-order in the 2D square plane above $T_{N}$.
In the presence of magnetic fields, the phase transition temperature decreases with increasing fields and almost disappears above 4 T, as shown in Fig. 2(b). 

\begin{figure}[t]
\begin{center}
\includegraphics[width=16pc]{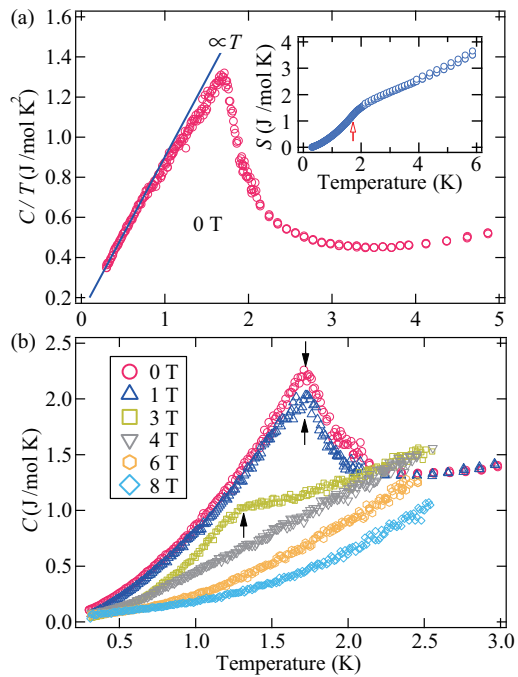}
\caption{(color online) Specific heat of ($o$-MePy-V)PF$_6$ at (a) zero-field and (b) various magnetic fields. The solid line shows the fit of $C/T$ $\propto$ $T$ below approximately 1 K. The inset shows the evaluated entropy at zero-field. The arrows indicate the phase transition temperatures. }\label{f2}
\end{center}
\end{figure}

\subsection{Magnetization}　
The inset of Fig.3 shows the temperature dependence of magnetic susceptibility ($\chi=M/H$) at 0.5 T . 
Above 50 K, $\chi$ follows the Curie-Weiss law, and the Weiss temperature is estimated to be ${\theta}_{\rm{w}}$ = $-$8.2(3) K. 
There is no distinct anomaly associated with the phase transition to the ordered state at $T_{N}$.
Below approximately 1 K, the magnetic susceptibility exhibits almost temperature-independent behaviour.
Because this low-temperature region is coincident with the region showing the $T^2$-dependent specific heat, the temperature-independent $\chi$ should reflect the ground-state property of the 2D frustrated system. 
The magnetization curve exhibits an unusually gradual increase up to approximately 10 T even below $T_{N}$ and then takes almost half the value of full saturation at fields between 10 and 25 T (1/2-plateau-like behaviour), as shown in Fig. 3.
At higher fields of approximately 30 T, the curve begins to increase again toward saturation.
Given the isotropic $g$ value of $\sim2.00$ in radical-based materials~\cite{a235Cl3V}, the saturation value of 0.97 $\mu _{\rm{B}}$/f.u. indicates a radical purity of 97${\%}$.
Note that the trend of the low-field gradual increase is completely different from the concave shape of the magnetization curve in general 2D quantum spin systems, in which quantum fluctuations are suppressed by applying magnetic fields, resulting in a rapid increase of magnetization~\cite{a235Cl3V, SrCuBO, tanaka, kagome}.

\begin{figure}[t]
\begin{center}
\includegraphics[width=16pc]{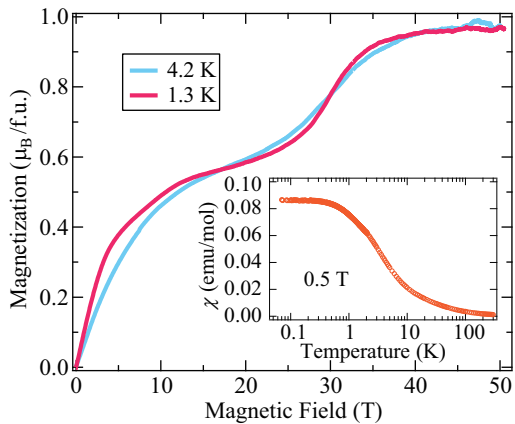}
\caption{(color online) Magnetization curve of ($o$-MePy-V)PF$_6$ at 1.3 K. The lower inset shows the temperature dependence of magnetic susceptibility ($\chi$ = $M/H$) at 0.5 T. }\label{f3}
\end{center}
\end{figure}

\subsection{Analysis by using the tensor network method}
Here, we examine the ground state of the expected $S$=1/2 Heisenberg frustrated square lattice by using the tensor network method.
Because multiple exchange interactions are evaluated from the $ab$ $initio$ calculations, it is difficult to determine the exact exchange parameters to reproduce the experimental behavior quantitatively.
However, in our previous studies, we confirmed that the $ab$ $initio$ MO calculations for verdazyl-based compounds provide reliable values of exchange interactions to qualitatively examine their intrinsic behavior~\cite{3Cl4FV, 2Cl6FV, b26Cl2V, pBrV,a235Cl3V}.
Thus, we assumed the evaluated exchange constants to examine the qualitative behavior and fixed the ratios as follows: $J_{2}/J_{1}=-0.82$, $J_{3}/J_{1}=-0.66$, $J_{4}/J_{1}=-0.61$, $J_{5}/J_{1}=0.26$, and $J_{6}/J_{1}=0.20$.
Figure 4(a) shows the calculated magnetization curve at $T=0$, which qualitatively reproduces the low-field convex function and subsequent 1/2-plateau-like behaviour. 
In the ground state at $H=0$, a collinear structure with twofold periodicity is realized in each site, as shown in Fig. 4(a).
Spins connected by the weakest FM $J_6$ in site 1 arrange in the opposite direction to minimize an increase in the ground-state energy due to the frustration. 
In the low-field region below $|H/J_{1}|{\simeq}0.8$, the spins in site 1 gradually tilt toward the field direction ($H//z$) with increasing field.
The average local magnetization for the field direction $\langle S_{z} \rangle$ in site 1 increases monotonically up to $|H/J_{1}|{\simeq}0.8$, but it is still not fully polarized because of the contribution of AFM $J_3$ and $J_4$ between two sites, as shown in Fig. 4(b).
Consequently, the intensively polarized spins in site 1 do not have sufficient degrees of freedom to modify the ground state, and site 2 forms an effective 1D $J_2$-$J_5$ chain.
This effective 1D chain induces quantum fluctuations attributed to its low-dimensionality, resulting in a field-enhanced reduction of the average local moment $(\langle S_{x}\rangle^{2}+\langle S_{y}\rangle^{2}+\langle S_{z}\rangle^{2})^{1/2}$ associated with the 1/2-plateau-like behaviour, as shown in Fig. 4(c).
In the field region above approximately $|H/J_{1}|{\simeq}1.2$, magnetizations in both sites increase toward the fully polarized phase.

\begin{figure}[t]
\begin{center}
\includegraphics[width=18pc]{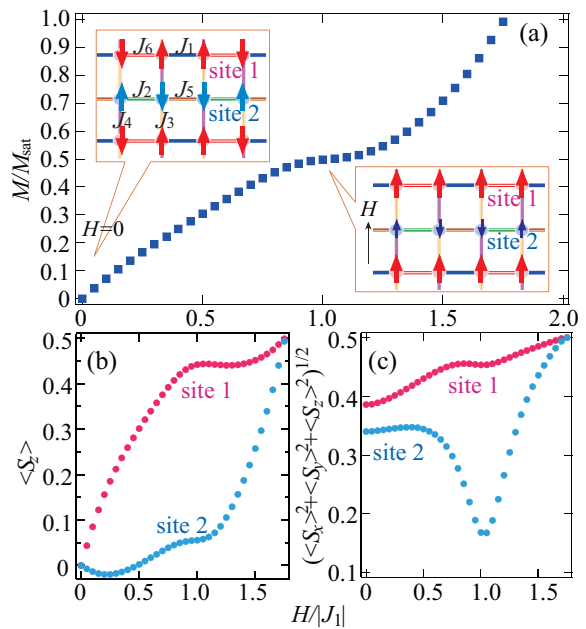}
\caption{(color online) The ground states under magnetic fields obtained from $D=6$ iTPS calculation. (a) Normalized magnetization curve, (b) average local magnetization, (c) average local moment at $T=0$ calculated using the tensor network method assuming the ratios of the evaluated exchange constants. The illustrations describe the predicted collinear spin structure at $H=0$ and field-enhanced reduction of the local moment near the 1/2-plateau-like phase.}\label{f4}
\end{center}
\end{figure}

\subsection{Nuclear magnetic resonance}
In order to investigate the local spin states, we performed $^{31}$P-NMR measurements.
Figure 5(a) shows the $^{31}$P-NMR spectra at 1.4 K for various magnetic fields along the $b$-axis.
Two crystallographically independent P sites, P(1) and P(2), are located on site 1 and site 2, respectively, as shown in Fig. 1(b).
We observed corresponding two-peak signals and evaluated the magnetic shift as a local magnetic field $H_{\rm{loc}}$, as shown in Fig. 5(b). 
It is difficult to determine the hyperfine coupling constants for this compound because the two peaks overlap with each other above 20 K owing to the small hyperfine couplings. 
However, the right peak shows a relatively large $H_{\rm{loc}}$, while the $H_{\rm{loc}}$ of the left peak stays near 0 at 1.4 K, as shown in Fig. 5(b). 
This result is consistent with the expectation from the numerical analysis shown in Fig. 4(b), where $\langle S_{z} \rangle$ for site 1 shows large values, while $\langle S_{z} \rangle$ for site 2 is almost zero at lower fields. 
Therefore, the right and left peaks can be attributed to P(1) and P(2) sites, respectively. 
The right peak is broadened below 3 T owing to the proximity of the ordered phase, as shown in Fig. 5(a). 
The $H_{\rm{loc}}$ at P(1) gradually increases with increasing $H$ up to 3 T.
Above 3 T, the $H_{\rm{loc}}$ at P(1) shows almost field-independent behaviour, while $H_{\rm{loc}}$ at P(2) shows a small negative shift. 
This shift at P(2) is considered to be related to the small increase of $\langle S_{z}\rangle$ for site 2 combined with a negative hyperfine coupling.
Figure 5(c) shows the temperature dependence of the nuclear spin-lattice relaxation rate $T_1^{-1}$  for P(1) and P(2). 
In the low-field regime at 2.5 T, $T_1^{-1}$ increases with decreasing temperature, indicating a magnetic phase transition with a critical slowing down at low temperature. 
Conversely, in the high-field regime at 5 T, $T_1^{-1}$ decreases with decreasing temperature because of the disappearance of the ordered phase. 
For the above measurements, the stretch exponent $\beta$ has a relatively small value probably because of the insufficient separation between two sites and/or the inhomogeneous distribution of the internal field.

\begin{figure}[t]
\begin{center}
\includegraphics[width=20pc]{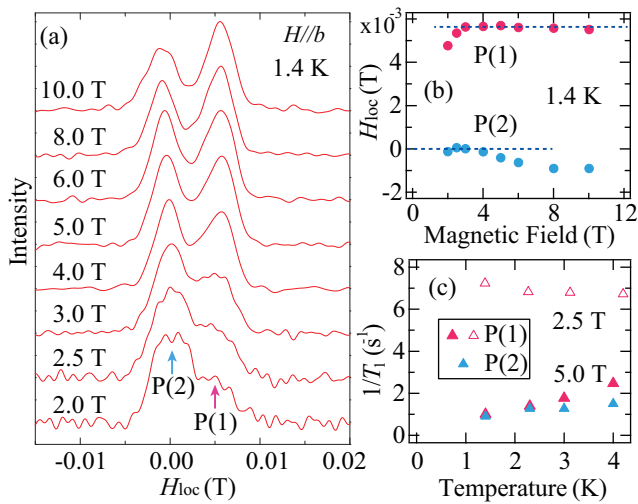}
\caption{(color online) (a) $^{31}$P-NMR spectra of ($o$-MePy-V)PF$_6$ at 1.4 K with various magnetic fields for $H//b$. (b) Local magnetic field $H_{\rm{loc}}$ of P(1) and P(2) evaluated from the spectra. (c) Temperature dependence of $T_1^{-1}$ for P(1) and P(2) at 2.5 and 5.0 T. }\label{f4}
\end{center}
\end{figure}

\section{Summary}
In summary, we succeeded in synthesizing single crystals of the verdazyl-based charge-transfer salt ($o$-MePy-V)PF$_6$. 
In the low-field regions below 4 T, a phase transition to the ordered state and a $T^2$ dependence of specific heat were observed, which indicate the formation of a 2D ground state with a linear dispersive mode.
Additionally, the low-temperature magnetization curve exhibited an unconventional gradual increase up to approximately 10 T and a subsequent 1/2-plateau-like behaviour at the fields between 10 and 25 T.
Our advanced numerical analysis using the tensor network method qualitatively demonstrated those magnetic behaviours with a field-induced effective 1D chain accompanied by field-enhanced quantum fluctuations.
Furthermore, the local magnetization and $T_1^{-1}$ probed by NMR measurements supported these findings.
Thus, we demonstrated that the new model compound of the $S$=1/2 frustrated square lattice, ($o$-MePy-V)PF$_6$, realizes field-enhanced quantum fluctuations, yielding unconventional quantum magnetism.
Our work revealed a new quantum nature of spins in magnetic materials and will open up new possibilities for material science focused on quantum physics.

\begin{acknowledgments}
This research was partly supported by Grant for Basic Science Research
Projects from KAKENHI (No. 15H03695, No. 15K05171, No. 15K17701, No. 15H03682, and
No. 17H04850), by the CASIO Science Promotion Foundation, by the Matsuda foundation, and by MEXT of Japan as a social and scientific priority issue (Creation of new functional devices and high-performance materials to support next-generation industries;CDMSI) to be tackled by using post-K computer. A part of this work was carried out at the Center for Advanced High Magnetic Field Science in Osaka University under the Visiting Researcher's Program of the Institute for Solid State Physics, the University of Tokyo, the Institute for Molecular Science, and on computers at the Super computer Center, ISSP, University of Tokyo.
\end{acknowledgments}



\begin{thebibliography}{99}
\bibitem{ITE_memo}
We have also performed the full update. However, no significant difference from the simple update has been observed.
\end{thebibliography}
\end{document}